\documentclass[a4paper]{llncs}
\usepackage{amsmath}
\usepackage{tikz}
\usepackage{amssymb}
\usepackage{pifont}
\usepackage{listings}
\usepackage{graphicx}
\usepackage{xspace}
\usepackage{pgf}
\usepackage[noend]{algpseudocode}
\usepackage{algorithm}
\usepackage{amsfonts}
\usepackage{amsmath}
\usepackage{pifont}
\usepackage{multicol}
\usepackage[justification=centering]{caption}
\usepackage{subcaption}
\usepackage{subfig}
\usepackage[framemethod=tikz]{mdframed}
\usepackage{framed}
\usepackage{wasysym,stmaryrd}
\usepackage{xcolor}
\usepackage{mathrsfs}
\usepackage{tikz}
\usepackage{galois}
\usetikzlibrary{arrows, automata, shapes, calc, positioning, matrix}
\usepackage{afterpage}

\algrenewcommand\algorithmicrequire{\textbf{Precondition:}}
\algrenewcommand\algorithmicensure{\textbf{Postcondition:}}

\newcommand{\tick}{\checkmark}
\newcommand{\Abs}[1]{{#1}^\sharp}

\newcommand{\path}{\mathit{path}}

\title{Danger Invariants}

\author{Cristina David \and Daniel Kroening \and Matt Lewis}
\institute{University of Oxford}

\newenvironment{keywords}{
       \list{}{\advance\topsep by0.35cm\relax\small
       \leftmargin=0cm
       \labelwidth=0.35cm
       \listparindent=0.35cm
       \itemindent\listparindent
       \rightmargin\leftmargin}\item[\hskip\labelsep
                                     \bfseries Keywords:]}
     {\endlist}

\makeatletter
\pgfdeclareshape{datastore}{
  \inheritsavedanchors[from=rectangle]
  \inheritanchorborder[from=rectangle]
  \inheritanchor[from=rectangle]{center}
  \inheritanchor[from=rectangle]{base}
  \inheritanchor[from=rectangle]{north}
  \inheritanchor[from=rectangle]{north east}
  \inheritanchor[from=rectangle]{east}
  \inheritanchor[from=rectangle]{south east}
  \inheritanchor[from=rectangle]{south}
  \inheritanchor[from=rectangle]{south west}
  \inheritanchor[from=rectangle]{west}
  \inheritanchor[from=rectangle]{north west}
  \backgroundpath{
    \southwest \pgf@xa=\pgf@x \pgf@ya=\pgf@y
    \northeast \pgf@xb=\pgf@x \pgf@yb=\pgf@y
    \pgfpathmoveto{\pgfpoint{\pgf@xa}{\pgf@ya}}
    \pgfpathlineto{\pgfpoint{\pgf@xb}{\pgf@ya}}
    \pgfpathmoveto{\pgfpoint{\pgf@xa}{\pgf@yb}}
\usepackage{tikz}
\usetikzlibrary{arrows, automata, shapes}

    \pgfpathlineto{\pgfpoint{\pgf@xb}{\pgf@yb}}
 }
}
\tikzset{>=stealth}
\makeatother

\begin{document}
\maketitle
\pagestyle{headings}  

\begin{abstract}
Static analysers search for overapproximating proofs of safety commonly
known as safety invariants.  Fundamentally, such analysers summarise traces
into sets of states, thus trading the ability to distinguish traces for
computational tractability.  Conversely, static bug finders (e.g.~Bounded
Model Checking) give evidence for the failure of an assertion in the form of
a counterexample, which can be inspected by the user.  However, static bug
finders fail to scale when analysing programs with bugs that require many
iterations of a loop as the computational effort grows exponentially with
the depth of the bug.  We propose a novel approach for finding bugs, which
delivers the performance of abstract interpretation together with the
concrete precision of BMC.  To do this, we introduce the concept of
\emph{danger invariants} -- the dual to safety invariants.
Danger invariants summarise sets of traces that are guaranteed to reach an error
state.  This summarisation allows us to find deep bugs without false alarms
and without explicitly unwinding loops.  We present a second-order formulation of
danger invariants and use the solver described in~\cite{kalashnikov} to
compute danger invariants for intricate programs taken from the literature.
\end{abstract}


\section{Introduction}


Safety analysers search for proofs of safety commonly known as {\em safety
invariants} by overapproximating the set of program states reached during
all program executions.  Fundamentally, they summarise traces into abstract
states, thus trading the ability to distinguish traces for computational
tractability~\cite{DBLP:conf/popl/CousotC77}.  Consequently, safety
analysers may generate bug reports that do not correspond to actual errors
in the code (i.e.~{\em false alarms}).  This is illustrated in
Figure~\ref{fig-overapp}.  False alarms are the primary barrier to the
adoption of static analysis technology outside academia.  Triage of true
errors and false alarms is a tedious and difficult task, and there are
reports that developers fare no better than coin
tossing~\cite{DBLP:conf/pldi/DilligDA12}.

\def\firstcircle{(-1,0) circle (1.4)}
\def\secondcircle{(2,0) circle (1)}
\def\thirdcircle{(-.5,0) circle (2)}
\begin{figure}
\begin{minipage}[b]{0.4\textwidth}
\centering
\begin{tikzpicture}[scale=0.6, every node/.style={scale=0.5}]
    \draw[thick, rounded corners] (-4,-2.5) rectangle (4,2.5);
    \draw (-2.2,2.3) node {A bug-free program};
    
    \begin{scope}[fill opacity=0.5]
      \fill[green] \firstcircle;
      \fill[red] \secondcircle;
    \end{scope}
  
    \fill[green, fill opacity=0.5] \thirdcircle;
    \draw[align=center] \firstcircle node {reachable\\ states};
    \draw[align=center] \secondcircle node {error\\ states};
    \draw[align=center] \thirdcircle node {}; 
    \begin{scope}
    \end{scope}

    \node[font=\bf] (F) at (1.5,-2.2) {false alarms};
    \draw[thick] (F) -- (1.3,0);
\end{tikzpicture} 
\caption{Generation of false alarms}
\label{fig-overapp}
\end{minipage}
\hfill
\begin{minipage}[b]{0.5\textwidth}
\centering
\def\firstcircle{(0,1.5) circle (2.3)}
\def\secondcircle{(0,-0.4) ellipse (.7 and .4)}
\begin{tikzpicture}[scale=0.6, every node/.style={scale=0.5}]
    \draw[thick, rounded corners] (-4,-1) rectangle (4,4);
    \draw (-2.2,3.8) node {A buggy program};
    
    \begin{scope}[fill opacity=0.3]
      \fill[green] \firstcircle;
      \fill[red] \secondcircle;
    \end{scope}
    
    \draw[align=center] \firstcircle node at (0,3.4) {reachable states};
    \draw[align=center] \secondcircle node {error\\ states};
  
  \draw [black] (0,2.8) ellipse ( 1cm and .3cm);
  \node[font=\bf] at (0,2.8) {A};
  \draw [black] plot [smooth] coordinates {(-1,2.8) (-.6,2.4) (0, 2.1) (.6,2.4) (1,2.8)};
  \node[font=\bf] at (0,2.3) {B};
  \draw [black] plot [smooth] coordinates {(-1,2.8) (-.6,2) (0, 1.6) (.6,2) (1,2.8)};
  \node[font=\bf] at (0,1.8) {C};
  \draw [black] plot [smooth] coordinates {(-1,2.8) (-.6,1.6) (0, 1.2) (.6,1.6) (1,2.8)};
  \node[font=\bf] at (0,1.4) {D};
  \draw [black] plot [smooth] coordinates {(-1,2.8) (-.6,1.2) (0, 0.8) (.6,1.2) (1,2.8)};
  \node[font=\bf] at (0,1) {E};
  \draw [black] plot [smooth] coordinates {(-1,2.8) (-.6,0.8) (0, 0.4) (.6,0.8) (1,2.8)};
  \node[font=\bf] at (0,.6) {F};
\end{tikzpicture}
\caption{Iterative unwinding of the transition relation} \label{fig-bmc}
\end{minipage}
\end{figure}

Conversely, static bug finders such as Bounded Model Checking (BMC) search for
proofs that safety can be violated.  Dually to safety proofs, we will call these
\emph{danger proofs}.  Static bug finders have the
attractive property that once an assertion fails, a counterexample trace is
returned, which can be inspected by the
user~\cite{DBLP:journals/fmsd/ClarkeBRZ01}.  The counterexample is thus the
proof that an assertion violation occurs.  In order to construct
such a danger proof, bounded model checkers compute underapproximations of the
reachable program states by progressively unwinding the transition relation. 
The downside of this approach is that static bug finders fail to scale when
analysing programs with bugs that require many iterations of a loop.  For
illustration, Figure~\ref{fig-bmc} depicts the successive unwinding of the
transition relation, where progressively larger sets of reachable program
states are labelled with letters from A to F.  The computational effort
required to discover an assertion violation (i.e.~to obtain an intersection
with the small ellipse labelled ``error states'') typically grows
exponentially with the depth of the bug.

Notably, the scalability problem is not limited to procedures that implement
BMC.  Approaches based on a combination of over- and underapproximations such
as predicate abstraction~\cite{DBLP:journals/toplas/ClarkeGL94} and lazy
abstraction with interpolants (LAwI)~\cite{DBLP:conf/cav/McMillan06} are not
optimised for finding deep bugs either.  The reason for this is that they can
only detect counterexamples with deep loops after the repeated refutation of
increasingly longer spurious counterexamples.
The analyser first considers a potential error trace with one loop
iteration, only to discover that this trace is infeasible.  Consequently,
the analyser increases the search depth, usually by considering one further
loop iteration.  This repeated unwinding suffers from the same exponential
blow-up as BMC.

\paragraph{Danger proofs.}

In this paper we propose a novel representation of a danger proof
based on trace summarisation.
%
%
We propose to merge the two core concepts of safety analysers based
on abstract interpretation and bug finders: trace summarisation
(for scalability purposes) and counterexample generation (for precision).
The intuition is that summarising traces is permissible as
long as the summary is guaranteed to contain at least one feasible
counterexample trace (i.e.  a trace that starts in an initial state and
reaches an error state).

The resulting summary is a dual of a safety invariant, which we refer to as a {\em danger invariant}.
As opposed to safety invariants, danger invariants do {\em not} necessarily include all the reachable program states,
but must contain at least one feasible execution trace. 
A danger invariant may encompass multiple paths through the program,
but contains enough information to directly read off a concrete error trace.
The danger invariant therefore amounts to a concise proof that such an
error trace exists.

From a practical point of view, danger invariants will allow the development of bug finding techniques that do not require explicit loop unwinding. 
We empirically show that danger invariants improve the scalability of bug finding, enabling the detection of deep bugs.
From a theoretical point of view, we propose a dual to over-approximating
safety invariants, which are at the core of safety analysis. 
Over-approximating invariants have received an enormous quantity of research
ever since the seminal work of Cousot and Cousot~\cite{DBLP:conf/popl/CousotC77}.
The equivalent of this body of work for invariants biased
towards bug finding is missing.

While the main technique that we use in this paper to compute danger
invariants is based on program synthesis (Section~\ref{sec:generation}),
given that the trace summarisation concept is common to both safety and
danger invariants, we also investigate how methods used for safety invariant
generation can be adapted for danger invariants
(Section~\ref{sec:fixed_point}).


\begin{mdframed}
\emph{Contributions:} 
\begin{itemize}
\item To the best of our knowledge, we propose the
first formulation of a {\em concise} proof of the existence of an error trace that allows trace summarisation
without false alarms.
\item We show that danger invariants improve the scalability of bug finding by using the second-order solver in \cite{kalashnikov} to infer 
danger invariants for programs with deep bugs.
\item We investigate how techniques used for inferring safety invariants can be adapted for danger invariants.
\item We generate danger invariants for a set of benchmarks taken from the literature.
\end{itemize}
\end{mdframed}


\section{From Counterexamples to Danger Invariants}

We represent a program $P$ as a transition system with state space $X$ and
transition relation $T \subseteq X \times X$.  For a state
$x \in X$ with $T(x,x')$, $x'$ is said to be a successor of $x$ under $T$.
We denote initial states by $I$ and error states by $E$.

\begin{definition}[Execution Trace]
\label{def:trace}
A program trace $\langle x_0 \ldots x_n \rangle$ is a (potentially infinite, in which case $n = \omega$) sequence of states,
such that any two successive states are related by the program's transition relation $T$, i.e.
\[\forall 0 \leq i < n . T(x_i, x_{i+1})\;.\]
\end{definition}

%

\begin{definition}[Counterexample]
A finite execution trace $\langle x_0 \ldots x_n \rangle$ is a
counterexample iff $x_0$ is an initial state, $x_0 \in I$, and $x_n$ is an
error state, $x_n \in E$.
\end{definition}

A counterexample is an instance of a danger proof: it provides evidence that an error state will be reached in some program execution. 
The question we try to answer in this paper is whether we can derive a more compact representation of a danger proof that does not
require us to explicitly write down every intermediate state.
Similarly to safety invariants, we obtain such a compact representation by summarising traces.
While this approach may involve overapproximation, the tricky part  
is retaining enough precision to ensure that a counterexample trace exists.
In the rest of the section, we will explain our formulation by starting from a safety invariant and  
progressively adjusting it to show the existence of a counterexample. 
For this purpose, we will refer to loops of the form $L(G, T, I, A)$ shown in Figure~\ref{fig:danger-2sat-loop}, which we encode 
with predicates for the initial states: $I(x)$, guard: $G(x)$, body: $T(x, x')$ and assertion: $A(x)$.

\begin{figure}
\begin{lstlisting}[mathescape=true,xleftmargin=.4\textwidth]
assume($I$);
while ($G$) $T$;
assert($A$);
\end{lstlisting}

\caption{The general form of a loop\label{fig:danger-2sat-loop}}
\end{figure}

\paragraph{Safety Invariants.}
Intuitively, a safety invariant is a set of states $S$ that includes every
state reachable via zero or more iterations of the loop, and which excludes
all error states.  More formally:

\begin{definition}[Safety Invariant]
\label{def:safety}
A predicate $S$ is a safety invariant for the loop $L(I, G, T, A)$ iff it
satisfies the following criteria:
\begin{align}
 \forall x . & I(x) \rightarrow S(x) \label{safety_base1} \\
 \forall x, x' . & S(x) \wedge G(x) \wedge T(x, x') \rightarrow S(x') \label{safety_inductive1} \\
 \forall x . & S(x) \wedge \neg G(x) \rightarrow A(x) \label{safety_safe1}
\end{align}
\end{definition}



\subsection{From Safety to Danger}

It is well known that Definition~\ref{def:safety} captures the notion of a
safety invariant, and that if a predicate $S$ exists that satisfies these
three criteria, then the loop $L$ is safe.  We will now consider what the
dual notion of a danger invariant might look like, and identify the criteria
defining it.

Let us begin by considering what happens if we take the natural step of
replacing criterion~\ref{safety_safe1} with its complement: if we exit the
loop in an $S$-state, we would like the assertion to fail.  This gives us
the following definition:

\begin{definition}[Doomed Loop Head]
\label{def:doom}
The head of the loop $L(I, G, T, A)$ is a \emph{doomed point}
\cite{doomstates} iff there exists a predicate $S'$ satisfying:
\begin{align}
 \forall x . & I(x) \rightarrow S'(x) \label{doom_base1} \\
 \forall x, x' . & S'(x) \wedge G(x) \wedge T(x, x') \rightarrow S'(x') \label{doom_inductive1} \\
 \forall x . & S'(x) \wedge \neg G(x) \rightarrow \lnot A(x) \label{doom_unsafe1}
\end{align}
\end{definition}

The term ``doomed program point'' was introduced in~\cite{doomstates}
and denotes a program location that, whenever reached, will inevitably lead
to an error regardless of the state in which it is reached. 
Definition~\ref{def:doom} applies this notion to the head of the loop $L$:
the predicate $S'$ provides proof that the head of the loop $L$ is a doomed
point, so if such an $S'$ exists then the loop will certainly fail. 
However, this definition is overly restrictive: in practice,
for most unsafe programs such an $S'$ does {\em not} exist.

For illustration, the program in Figure~\ref{fig:nodoom} is unsafe (any
execution starting with $x > 10$ will lead to the assertion failing) but
since there are \emph{some} initial states that do not lead to an error
(i.e.~every state with $x \leq 10$) there are no doomed points.  Thus, we
weaken Definition~\ref{def:doom} by introducing {\em doomed
program states}: we weaken criterion~\ref{safety_base1} to say that there
must be \emph{some} initial state leading to an error rather than
\emph{every} initial state.  This gives us the following:

\begin{definition}[Doomed Program State]
\label{def:doomstate}
There is trace containing an error state, starting from the head of the loop $L(I, G, T, A)$ if a predicate $S''$ exists
satisfying:
\begin{align}
 \exists x_0 . & I(x_0) \rightarrow S''(x_0) \label{doomstate_base1} \\
 \forall x, x' . & S''(x) \wedge G(x) \wedge T(x, x') \rightarrow S''(x') \label{doomstate_inductive1} \\
 \forall x . & S''(x) \wedge \neg G(x) \rightarrow \lnot A(x) \label{doomstate_unsafe1}
\end{align}
\end{definition}

\begin{figure}

\begin{tabular}{ccc}
\begin{subfigure}[t]{.3\textwidth}
\begin{lstlisting}[basicstyle=\scriptsize]
x = *;
while (x < 10) {
    x++;
}

assert (x == 10);
\end{lstlisting}
\caption{An unsafe loop with no doomed program points.\label{fig:nodoom}}
\end{subfigure}

&

\begin{subfigure}[t]{.3\textwidth}
\begin{lstlisting}[basicstyle=\scriptsize]
x = 0;
while (x < 10) {
    if(*) break;
    x++;
}
assert (x == 10);
\end{lstlisting}
\caption{An unsafe loop with no doomed states.\label{fig:nodoomstates}}
\end{subfigure}

&

\begin{subfigure}[t]{.3\textwidth}
 \begin{lstlisting}[basicstyle=\scriptsize]
  x = 0;
  y = 0;
  while (x < 10) {
    y++;
  }
  assert(x < 10);
 \end{lstlisting}
 \caption{A safe program with no finite error traces.\label{fig:nonterm}}
\end{subfigure}

\end{tabular}
\caption{Illustrative programs -- * means nondeterministic choice.\label{fig:counterexamples}}
\end{figure}

Definition~\ref{def:doomstate} weakens the notion of a doomed location. 
Rather than requiring every trace including the doomed location to reach an
error, we only require that there is some doomed initial state, i.e.  every
trace including that \emph{state} will reach an error.  However, this is
still too strong a condition!  Figure~\ref{fig:nodoomstates} shows an unsafe
program which has no doomed states -- every state in the loop has some
successor that leads to a state in which the assertion holds (i.e.~every
trace in which the \lstinline|break| is never executed).  To work around
this, we can weaken criterion~\ref{safety_inductive1} to say that each state
must have \emph{some} successor leading to an error.

\begin{definition}[Partial Danger]
\label{def:partialdanger}
There is a trace containing an error state, starting from the head of the loop $L(I, G, T, A)$ if a predicate $S'''$ exists
satisfying:
\begin{align}
 \exists x_0 . & I(x_0) \rightarrow S'''(x_0) \label{partial_base1} \\
 \forall x . & S'''(x) \wedge G(x) \rightarrow \exists x' . \wedge T(x, x') \wedge S'''(x') \label{partial_inductive1} \\
 \forall x . & S'''(x) \wedge \neg G(x) \rightarrow \lnot A(x) \label{partial_unsafe1}
\end{align}
\end{definition}

We are very nearly done: Definition~\ref{def:partialdanger} captures
that there is some trace containing an error state starting from some
initial state.  However, our definition of an execution trace
(Definition~\ref{def:trace}) includes infinite traces.  Thus, the trace
containing the error may be infinite and the error state will not be
reachable at all.  For example, consider Figure~\ref{fig:nonterm}.  A
possible partial danger invariant is `\lstinline|true|', which meets all of the
criteria \ref{partial_base1}, \ref{partial_inductive1} and
\ref{partial_unsafe1}.  However, the program is in fact safe -- it contains
no terminating traces and so the assertion is never even reached. 
Definition~\ref{def:partialdanger} captures \emph{partial danger},
which disregards the termination behaviour of the program.  Instead, what we
really want is \emph{total danger}, which will guarantee that there is some
finite trace culminating in an error.  To ensure that the error traces are
finite, we will introduce a \emph{ranking function}, which will serve as a
proof of termination.  Below we recall the definition of a ranking function:

\begin{definition}[Ranking function]
A function ${R:X\to Y}$ is a \emph{ranking function} for the
transition relation $T$ if $Y$ is a well-founded set with order $>$ and 
$R$ is injective and monotonically decreasing with respect to $T$.  That is
to say:
$$\forall x, x' \in X. T(x, x') \Rightarrow R(x) > R(x')$$
\end{definition}

This is the final piece we need to define danger invariants:

\begin{definition}[Danger Invariant]
 \label{def:danger}
 A pair $\langle D, R \rangle$ of a predicate and a ranking function is a danger invariant for the loop
 $L(I, G, T, A)$ iff it satisfies the following criteria:
 \begin{align}
 \exists x_0 . & I(x_0) \wedge D(x_0) \label{danger_base} \\
 \forall x . & D(x) \wedge G(x) \rightarrow R(x) > 0  \wedge \exists x' . T(x, x') \wedge D(x') \wedge R(x') < R(x) \label{danger_inductive} \\
 \forall x . & D(x) \wedge \neg G(x) \rightarrow \neg A(x) \label{danger_danger}
\end{align}
\end{definition}

\begin{theorem}[Danger Invariants Prove Bugs]
 The loop $L(I, G, T, A)$ is unsafe iff there exists a danger invariant satisfying the criteria in Definition~\ref{def:danger}.
 In other words, the existence of a danger invariant is a necessary and sufficient condition for the reachability of a bug.
\end{theorem}

\section{Second-Order Formulation of Danger Invariants}

The problem of program verification can be reduced to the problem of finding
solutions to a second-order
constraint~\cite{DBLP:conf/pldi/GrebenshchikovLPR12,DBLP:conf/pldi/GulwaniSV08}. 
In order to give a second-order formulation of a danger invariant, we will use a  
fragment of second-order logic decidable over finite domains that we defined in \cite{kalashnikov}, and to whose satisfiability problem we refer as Second-Order SAT. 
Next, we recall this fragment:

\begin{definition}[Second-Order SAT]
\label{def:2sat}
 \[
  \exists S_1 \ldots S_m . Q_1 x_1 \ldots Q_n x_n . \sigma
 \]
 Where the $S_i$ range over predicates, 
the $Q_i$ are either $\exists$ or $\forall$,
$x_i$ range over Boolean values and $\sigma$ is a quantifier-free propositional formula
that may refer to both the first-order variables $x_i$ and the second-order variables $S_i$. 
\end{definition}

\begin{figure}
\begin{framed}
\begin{definition}[Danger Invariant Formula {\bf [DI]}]
\label{def:di}
\begin{align*}
 \exists D, R, x_0 . \forall x . \exists x' . & I(x_0) \wedge D(x_0) ~ \wedge \\
 										  & D(x) \wedge G(x) \rightarrow T(x ,x') \wedge D(x') ~ \wedge \\
                                                                                  & R(x) > 0 \wedge R(x) > R(x')~ \wedge \\
 										  & D(x) \wedge \neg G(x) \rightarrow \neg A(x)
\end{align*}
\end{definition}

\begin{definition}[Skolemized Danger Invariant Formula {\bf [SDI]}]
\label{def:sdi}
\begin{align*}
 \exists D, R, N, x_0 . \forall x . & I(x_0 \wedge D(x_0) ~ \wedge \\
 							     & D(x) \wedge G(x) \rightarrow R(x) > 0 \wedge T(x, N(x)) \wedge D(N(x)) \wedge R(x) > R(N(x))\\
 								 & D(x) \wedge \neg G(x) \rightarrow \neg A(x)
\end{align*}
\end{definition}

\end{framed}
\caption{Existence of a danger invariant as second-order SAT\label{fig:danger_formulae}}
\end{figure}

Our first second-order formulation of a danger invariant  
is captured in the second-order SAT formula {\bf [DI]} of Definition~\ref{def:di}.  
In this definition, in order to specify that from each $D$-state we can reach another by iterating the loop
once, we require quantifier alternation over the first order variables. However, the solver from \cite{kalashnikov}
that we want to use requires eliminating the extra level of quantifier alternation (the inner existential quantifier) by using Skolem functions.

If the transition relation $T$ is deterministic, then we do not need the
quantifier alternation, since each $x$ has exactly one successor $x'$. 
Thus, we can just replace the inner $\exists x'$ in the formula {\bf [DI]}
by $\forall x'$.  However, if $T$ is non-deterministic, we must find a
Skolem function $N$ which resolves the non-determinism by telling us exactly
which successor is to be chosen on each iteration of the loop.  This is
shown in the formula {\bf [SDI]} of Definition~\ref{def:sdi}.




\subsection{Danger Invariants Generation} \label{sec:generation}

In this section, we discuss how to solve the constraints generated for
danger invariants.
%
%
%
In~\cite{kalashnikov}, we show that Second-Order SAT is polynomial-time reducible to finite synthesis, and 
finite-state program synthesis is NEXPTIME-complete.
%
Next, we provide a short description of the 
program synthesis algorithm and, for more details, we direct the reader to \cite{kalashnikov}.
Our algorithm is sound and complete for the
finite-state synthesis decision problem.  In the case that a specification
is satisfiable, our algorithm produces a {\em minimal} satisfying
program. We use Counterexample Guided Inductive Synthesis
(CEGIS)~\cite{lezama-thesis,sketch} to find a
program satisfying our specification.

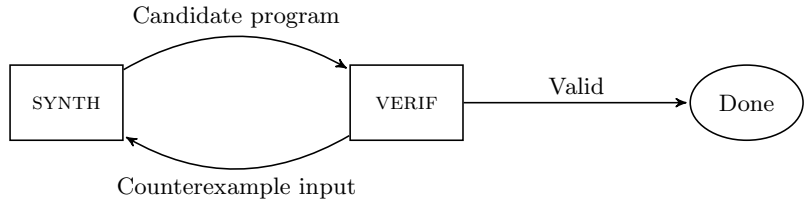
\begin{figure}
 \centering
 \begin{tikzpicture}[scale=0.5,->,>=stealth',shorten >=1pt,auto,
 semithick, initial text=]

  \matrix[nodes={draw, fill=none, scale=1, shape=rectangle, minimum height=1cm, minimum width=1.5cm},
          row sep=2cm, column sep=3cm] {
   \node (synth) {\sc synth};
   &
   \node (verif) {\sc verif}; 
   &
   \node[ellipse] (done) {Done}; \\
  };

   \path
    (synth) edge [bend left] node {Candidate program} (verif)
    (verif) edge [bend left] node {Counterexample input} (synth)
    (verif) edge node {Valid} (done);
 \end{tikzpicture}
 
 \caption{Abstract synthesis refinement loop
 \label{fig:abstract-refinement}}
\end{figure}

As illustrated in Figure~\ref{fig:abstract-refinement}, the algorithm is
divided into two procedures: {\sc synth} and {\sc verif}, which interact via
a finite set of test vectors {\sc inputs}.
By using explicit proof search, symbolic bounded model checking and genetic
programming with incremental
evolution~\cite{langdon:fogp,brameier2007linear}, the {\sc synth} procedure
tries to find an existential witness $P$ that satisfies the partial
specification:
\[
 \exists P . \forall x \in \text{\sc inputs} . \sigma(x, P)
\]

If {\sc synth} succeeds in finding a witness $P$, this witness is a
candidate solution to the full synthesis formula.  We pass this candidate
solution to {\sc verif} which determines whether it does satisfy
the specification on all inputs by checking satisfiability of the
verification formula:
\[
 \exists x . \lnot \sigma(x, P)
\]

If this formula is unsatisfiable, the candidate solution is in fact a
solution to the synthesis formula and so the algorithm terminates. 
Otherwise, the witness $x$ is an input on which the candidate solution fails
to meet the specification.  This witness $x$ is added to the {\sc inputs}
set and the loop iterates again.  


\subsection{Generalised Safety-Danger Formula}

The decision procedure introduced in \cite{kalashnikov} and briefly described above  
relies on a small-model argument to determine that a formula is unsatisfiable and, in practice, 
it is usually unable to prove unsatisfiability.  
Therefore we would like to only provide satisfiable formulae
whenever possible.  Since the program we are analysing is either safe or unsafe, 
and assuming that a proof is expressible in our logic, 
a program either accepts a safety invariant or a danger invariant. 
We model this as a disjunction 
in the formula {\bf [GS]} of Definition~\ref{def:gs}.  {\bf [GS]} is a theorem of second-order logic, and 
our decision procedure will always be able to find witnesses $S, D, N, R, y_0$ demonstrating its truth, provided such a witness
is expressible in our logic.  The synthesised
predicate $S$ is a purported safety invariant and the $D, N, R, y_0$ constitute a purported danger invariant.
If $S$ is really a safety invariant, the program is safe, otherwise
$D, R$ (with witnesses to the existence of an error trace with Skolem function $N$ and initial state $y_0$) will be a danger invariant and the program is unsafe.
Exactly one of these proofs will be valid, i.e. either $S$ will satisfy the criteria for a safety invariant, or $D, N, R, y_0$ will satisfy
the criteria for a danger invariant.  We can simply check both cases and discard whichever ``proof'' is incorrect.

\begin{figure}
\begin{framed}
\begin{definition}[Generalised Safety Formula {\bf [GS]}]
\label{def:gs}

\begin{align*}
\exists S, D, N, R, y_0 . \forall x, x', y  . & \left(\begin{aligned}%
						  & I(x) \rightarrow S(x) ~ \wedge \\
						  & S(x) \wedge G(x) \wedge T(x, x') \rightarrow S(x') ~ \wedge \\
						  & S(x) \wedge \neg G(x) \rightarrow A(x) \\
                                               \end{aligned}\right) ~ \vee \\
					   &   \left(\begin{aligned}%
					          & I(y_0) \wedge D(y_0) ~ \wedge \\
					          & D(y) \wedge G(y) \rightarrow R(y) > 0 \wedge T(y, N(y)) \wedge D(N(y)) \\
                                                  & \qquad \qquad \qquad \wedge R(y) > R(N(y)) \wedge \\
					          & D(y) \wedge \neg G(y) \rightarrow \neg A(y) \\
					         \end{aligned}\right)
\end{align*}

\end{definition}

\end{framed}
\caption{General second-order SAT formula characterising safety\label{fig:general_safety_formulae}}
\end{figure}

\section{Illustrative Examples}

To illustrate the use of danger invariants for compactly representing danger
proofs and finding deep bugs, as well as the shortcomings of the existing
bug finding techniques, we consider the programs in
Figure~\ref{fig:danger-motivation}, where (a), (b), (c), (d) contain deep
bugs, whereas (e) models a buffer overflow.

For program (a), any execution trace violates the assertion unless the
nondeterministic choice (denoted by ``*'') is such that $y$ is incremented
exactly 999999 times out of the 1000000 iterations of the loop.  In
order to analyse this program bounded model checkers have to completely
unwind the loop.  The resulting SAT instance is very large, so large in fact
that solving it will almost certainly take far too long to be practical. 
Hybrid approaches such as predicate abstraction and LAwI will have to
progressively unwind the loop in order to refute spurious counterexamples of
increasing length.  Again, this is most likely to take too long to be
practical.

In our case, in order to prove that the assertion can be violated, we need
to find a danger invariant.  This means that we need an approximation of the
set of states reachable during the program's execution that must include a
feasible counterexample trace.  Of course there may be several such
invariants.  One possibility for this example is $D(x,y) = x < y$ and
ranking function $R(x, y) = -x$.  $D$ holds in the initial state where $x=0$
and $y=1$, and it is inductive with respective to the loop's body if the
nondeterministic choice is given by the Skolem function $N_y(x, y) = y+1$. 
That is: 
\[\forall x,y,x'. x < y \rightarrow x'=x+1 \wedge x' < N_y(x,y)\]

Program (b) is a version of (a) with additional nondeterminism. Now, in each
loop iteration, $x$ may or may not be incremented.  This modification
substantially increases the number of reachable program states, as well as
introducing a potential non-terminating behaviour as $x$ may not reach
1000000.  As a result, bounded model checkers will loop forever trying to
generate the SAT instance corresponding to the unwound loop.  Similarly,
predicate abstraction and LAwI based approaches are even less likely to be
practical than for version (a) of the program due to the increase in the
number of reachable states.  We synthesise the same $D$
as for program (a) as it still contains a feasible counterexample trace
regardless of the additional nondeterminism.  Here, termination depends on
the Skolem function giving us the successor of $x$.  Thus, we can have
$R(x,y) = -x$ and $N_x(x, y) = x+1$.

Program (c) is similar to (a), with the exception that the assertion
is now negated.  This example is more intricate as the danger invariant
needs to capture the evolution of $x$ and $y$ from the the initial state
where they are not equal to a final state where there are (and hence they
cause the assertion to fail).  One such invariant is $D(x,y) = y == (x
<1?  1:x)$ and $R(x, y) = -x$.  Essentially, this invariant says that
$y$ must not be incremented for the first iteration of the loop (until
$x$ reaches the value 1), and from that point, for the rest of the
iterations, $y$ gets always incremented such that $x == y$.  For this case, $D$
is a compact and elegant representation of exactly one feasible
counterexample trace.  The witness Skolem function that we get is $N_y(x,y)
= (x<1?y:y+1)$.




Program (d) is taken from \cite{DBLP:conf/sigsoft/GulavaniHKNR06}, and it is meant to illustrate 
the difficulty faced by tools based on predicate abstraction and abstraction refinement 
with deterministic loops with a fixed execution 
count as many iterations of the iterative refinement algorithm correspond to spurious executions of the loop body. 
The assertion fails, but proving this requires 
1000000 iterations of the refinement loop, resulting in the introduction of the
predicates ($i==0$),($i==1$),...,($i==1000000$) one by one.  For us, some of
the possible danger invariants are $D(i,c,a) = a{\leq}-10$, $D(i,c, a) =
a==0$, $D(i,c,a) = a{\leq}0$ and $R(i,c,a) = true$.

Program (e) models a check for a
buffer overflow.  The buffer overflow does happen whenever $x$ is big enough
such that the computation of $x * 4$ overflows.  We find the danger
invariant $D(x, i, len) = (i==0 \wedge len==0 \wedge x \neq 0)$ and ranking
function $R(x, i, len) = true$.  This basically says that the computation of
$len$ overflowed resulting in $len = 0$ (while $x \neq 0$).  Consequently,
the loop is not taken such that $i$ stays $0$ and the ranking function is
$true$.  Note that this example has the assertion inside the loop, and
required some trivial preprocessing in order to move it outside the loop.

\begin{figure}[H]
\centering
\begin{tabular}{cc}
\begin{subfigure}[b]{0.5\textwidth}
\begin{lstlisting}
x = 0; y = 1;
while (x < 1000000) {
    x++;
    if(*) y++;
}

assert ( x == y ) ;
\end{lstlisting}
\caption{}
 \label{fig:danger-motivation.a}

\hfill
\begin{lstlisting}

x = 0; y = 1;
while (x < 1000000) {
    if(*) x++;
    if(*) y++;
}

assert ( x == y ) ;
\end{lstlisting}
\caption{}
 \label{fig:danger-motivation.b}

\hfill

\begin{lstlisting}
x = 0; y = 1;
while (x < 1000000) {
    x++;
    if(*) y++;
}

assert ( x != y ) ;
\end{lstlisting}
\caption{}
 \label{fig:danger-motivation.c}
\end{subfigure}%

&

\begin{subfigure}[b]{0.5\textwidth}

\begin{lstlisting}

void foo(int a)
{
 int i, c;
 i = 0;
 c = 0;
 while (i < 1000000) {
   c=c+i;
   i=i+1;
 }
 assert(a > 0);
}

\end{lstlisting}
\caption{Adapted from \cite{DBLP:conf/sigsoft/GulavaniHKNR06}}
 \label{fig:danger-motivation.d}

\hfill

\begin{lstlisting}
int main(void) {
  unsigned int x, i, len;

  len = x * 4;

  for (i = 0; i < x; i++) {
    assert(i * 4 < len)
  }
}
\end{lstlisting}
\caption{Checking for a buffer overflow caused by an integer overflow.}
 \label{fig:danger-motivation.e}
\hfill

\end{subfigure}






\end{tabular}
\caption{Motivational examples.\label{fig:danger-motivation}}
\end{figure}

\section{Danger Invariants in Relation to Other Formalisms} \label{sec:fixed_point}

In this section, we relate danger invariants to other existing formalisms. 
For this purpose, we start by providing a characterisation of danger
invariants as a fixed point computation.  First, we define the set of
program executions starting in an initial state, $E_\mathit{fwd}$, and the set of
program executions ending in an error state, $E_{bck}$.  The set of program
execution traces starting in an initial state (real executions) is
$$T_\mathit{fwd} = \{\langle x_0 ... x_n \rangle ~|~ \forall i.~ T(x_i, x_{i+1}) \wedge x_0 \in I\} = \mathit{lfp}(F_\mathit{fwd})$$ 
where $F_\mathit{fwd}$ constructs execution traces by taking a transition forward:
$$F_\mathit{fwd} = \lambda S. \{\langle x \rangle ~|~ x \in I\} \cup \{\langle x_0 ... x_n, x_{n+1} \rangle ~|~ \langle x_0 ... x_n \rangle \in S \wedge T(x_n, x_{n+1})\}$$
The set of program execution traces ending in an error state is 
$$T_{bck} = \{\langle x_0 ... x_n \rangle ~|~ \forall i.~ T(x_i, x_{i+1}) \wedge x_n \in E\} = \mathit{lfp}(F_{bck})$$
where $F_{bck}$ constructs execution traces by taking a transition backward:
$$F_{bck} = \lambda S. \{\langle x \rangle ~|~ x \in E \} \cup \{\langle x_{-1}, x_0 ... x_n \rangle ~|~ \langle x_0 ... x_n \rangle \in S \wedge T(x_{-1}, x_{0})\}$$

Similar to~\cite{DBLP:conf/sas/Rival05}, we can now express the set of
execution traces starting in an initial state and ending in an error state
as $Err = \mathit{lfp}(F_\mathit{fwd}) \cap \mathit{lfp}(F_\mathit{bck})$.  Then, a danger invariant
is an approximation of $\mathit{lfp}(F_\mathit{fwd})$ (either $ D \supseteq
\mathit{lfp}(F_\mathit{fwd})$ or $ D \subseteq \mathit{lfp}(F_\mathit{fwd})$), such that it
contains at least one error trace $D \cap Err \neq \emptyset$.

\subsection{Abstract Interpretation}

Given that direct computation of the set of possible execution traces of a program is most of the times infeasible, 
an abstract interpretation based analysis conservatively computes at each program point a set of abstract states representing 
an overapproximation of the possible concrete program states \cite{DBLP:conf/popl/CousotC77}.   
The analysis assigns to each program location an abstract value from an
abstract domain. 
The concretisation mapping $\gamma$ is defined such that 
for every program location $l$ produces an abstract state 
$x^{\#}$, such that  
$\gamma(x^{\#})$ contains all the concrete states reachable at location $l$.



The abstract forward interpreter $\Abs{F_\mathit{fwd}}$ is inherently overapproximating,\\
$\mathit{lfp}(F_\mathit{fwd}) \subseteq \gamma(\Abs{\mathit{lfp}}(\Abs{F_\mathit{fwd}}))$.
This means that, whenever an error is signalled at program location $l$, the
alarm may be spurious.  In order to decide whether an alarm is genuine,
abstract interpretation based techniques require backward analysis starting
from the error state.  Various solutions have been already proposed in the
abstract interpretation
literature~\cite{DBLP:conf/nfm/BrauerS12,erez-thesis,DBLP:conf/sas/Rival05}. 
Next, we see which of these can be used to compute danger invariants.

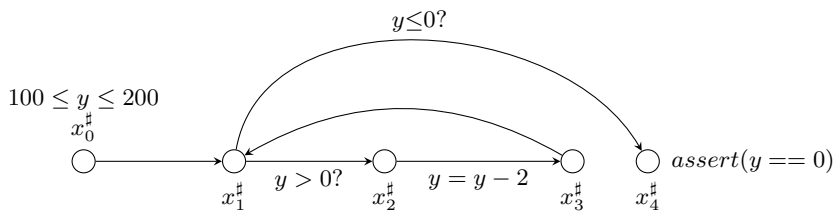
\begin{figure}
\begin{tikzpicture}[->]

  \node[circle, draw, label={[align=center]$100 \leq y \leq 200$\\$\Abs{x}_0  $}] (x0) at (4,0){};
  \node[circle, draw, right of=x0] (x1) at (5,0) [label=below:$\Abs{x}_1$] {};
  \node[circle, draw, right of=x1] (x2) at (7,0) [label=below:$\Abs{x}_2$] {};
  \node[circle, draw, right of=x2] (x3) at (9.5,0) [label=below:$\Abs{x}_3$] {};
  \node[circle, draw, right of=x3] (x4) [label=below:$\Abs{x}_4$] {};
  \node[right of=x4] (x5) at (11.9,0) {$assert(y==0)$};

  \draw (x0) to node[below] {} (x1);
  \draw (x1) to node[below] {$y>0?$} (x2);
  \draw (x2) to node[below] {$y=y-2$} (x3);
  \draw (x3) edge[out=150,in=30] node {} (x1); 
  \draw (x1) edge[out=80,in=120] node[above] {$y{\leq}0?$} (x4);
 
\end{tikzpicture}
\caption{The CFG of an erroneous program: the labels on the vertices denote the corresponding abstract program states and the arcs correspond to instructions in the program.}
\label{fig:ex_cfg}
\end{figure}

For illustration purposes, we will use the program in Figure~\ref{fig:ex_cfg} (adapted from \cite{DBLP:conf/nfm/BrauerS12}) as a running example throughout this section. 
The program takes an input variable $y$ bounded between 100 and 200 and iteratively decreases it by 2.
The program is erroneous as the assertion $y == 0$ is violated whenever the initial value of $y$ is odd.
The abstract states $\Abs{x_0}$ to $\Abs{x_4}$ inferred by a forward analysis based on the 
interval domain are listed next:
$\Abs{x_0} = [100,200]$,
$\Abs{x_1} = [-1,200]$,
$\Abs{x_2} = [1,200]$,
$\Abs{x_3} = [-1,198]$,
$\Abs{x_4} = [-1,0]$.
   
Since $\Abs{x_4}$ violates the assertion $y==0$, in order to check whether it is genuine, 
the negation of the assertion $y {<} 0 \wedge y{>}0$ must be propagated backwards.
Given that the result of the forward analysis is a sound overapproximation, the negation of the assertion can be intersected with $\Abs{x_4}$ 
resulting in $y = -1$, which is then propagated backwards. If this propagation results in the set of states 
being empty at any program location, then the error is a false alarm. Otherwise, it is genuine. 
The main challenge is the presence of loops. Given a state after a loop, it is non-trivial to infer a state that
is valid prior to entering the loop. In particular, it is necessary to assess how
often the loop body needs to be executed to reach the exit state. 


In \cite{DBLP:conf/nfm/BrauerS12} Brauer and Simon use an overapproximating affine analysis to estimate the number of loop iterations. 
Thus, they are able to obtain $y = 2n-1 \wedge n \in [1, 63]$ at location $l_1$ (corresponding to $\Abs{x_1}$ in the CFG). 
In order to obtain a danger invariant from this, we just need to add the error state $y=-1$. Thus, we obtain $D(y) = (y == 2n-1 \wedge n \in [0, 63])$, which is 
is guaranteed to contain a concrete counterexample. 
In \cite{erez-thesis}, Erez performs a bounded search for backward traces up to a given depth. For the given example,
the first counterexample found is: $\langle y=-1, y=1, ..., y = 101\rangle$.
Thus, a corresponding danger invariant is $D(y) = (y==-1 \vee y==1 \vee ... \vee y==101)$, or $D(y) = (y == 2n-1 \wedge n \in [0, 51])$.
While the results of both these techniques can be immediately used to compute danger invariants, this is not true for other works on eliminating false alarms in abstract interpretation
based techniques such as \cite{DBLP:conf/sas/Rival05}, where Rival computes an overapproximation of $Err$ by using in his backward analysis the same domains 
as in forward analysis. Thus, the result is not a danger invariant and cannot ensure the existence of a true error.

\subsection{Bounded Model Checking}
For a loop $L(I, G, T, A)$, a bounded model checker progressively unwinds the transition relation up to a depth $k$. As such:

$$T_\mathit{fwd}^k = \mathit{lfp}(F_\mathit{fwd}^k)$$ 
where $F_\mathit{fwd}^k$ constructs execution traces by taking a transition forward:
$$F_\mathit{fwd}^k = \lambda S. \{\langle x \rangle ~|~ x \in I\} \cup \{\langle x_0 ... x_n, x_{n+1} \rangle ~|~ \langle x_0 ... x_n \rangle \in S \wedge T(x_n, x_{n+1}) \wedge n{<}k\}$$

Thus, BMC finds counterexample traces of the form $\langle x_0, ..., x_n \rangle$, where $x_0 \in I$,
$x_n \in E$ and $n{\leq}k$. Such a counterexample directly corresponds to the danger invariant $D(x) = \bigvee_{i=0,n} x_i$.
For the running example, we have used CBMC~\cite{ckl2004} and obtained
the counterexample trace $\langle y=101, y=99, ...  ,y=-1 \rangle$, which
corresponds to the danger invariant $D(y) = (y == 101 \vee y==99 \vee \ldots
\vee y==-1)$.


\subsection{Linear Invariants}

There is a lot of work on the generation of linear invariants of the form
$c_1x_1 + \ldots + c_nd_n + d \leq 0$
\cite{DBLP:conf/cav/ColonSS03,DBLP:conf/cav/0001A14}.  The main idea behind
these techniques is to treat the coefficients $c_1,\ldots, c_n, d$ as
unknowns and generate constraints on them such that any solution corresponds
to a safety invariant.  In~\cite{DBLP:conf/cav/0001A14}, Colon et
al.~present a method based on Farkas' Lemma, which synthesises linear
invariants by extracting non-linear constraints on the coefficients of a
target invariant from a program.  In a different work, Sharma and Aiken use
randomised search to find the coefficients~\cite{DBLP:conf/cav/0001A14}.  It
would be interesting to investigate how these methods can be adapted for
generating constraints on the coefficients $c_1,\ldots, c_n, d$ such that
solutions correspond to linear danger invariants.

\section{Experimental Results}
To evaluate our algorithm, we implemented the {\sc Dangerzone} tool, which generates a
danger specification from a C program and calls the second-order SAT
solver discussed in~\cite{kalashnikov} to obtain a proof.  We ran the
resulting prover on 20 buggy programs including the running examples in the paper, 
some examples from the literature and
some from SV-COMP'15~\cite{svcomp15}.  Our benchmarks do not make use of arrays or
recursion.  We do not have arrays in our logic and we had not implemented
recursion in our frontend (although the latter can be syntactically
rewritten to our input format).

For each benchmark we infer a danger invariant, a ranking function, an initial state and Skolem functions 
witnessing the nondeterminism. 
To provide a comparison point, we also ran {\sc CBMC}~\cite{ckl2004} on
the same benchmarks.  
For {\sc CBMC}, we manually provided sufficient unwinding limits for each program. 
Each tool was given a time limit of 1800\,s, and was
run on an unloaded 4-core 3.30\,GHz i5-2500k with 8\,GB of RAM.  The
results of these experiments are given in Figure~\ref{fig:experiments}.

On programs with shallow bugs (i.e. bugs that can be reached after a small number of loop iterations), {\sc CBMC} is much faster than {\sc Dangerzone}.
However, {\sc Dangerzone} performs much better on programs with deep bugs providing empirical evidence that danger invariants 
improve the scalability of static bug finding.
We feel that the performance difference for shallow bugs 
is inherent to the difference in the two approaches -- our solver is more general (and less engineered)
than {\sc CBMC}, in that it provides a complete proof system for both
danger and safety. The two examples on which {\sc Dangerzone} times out 
have an intricate control flow structure which requires specific constants to be inferred
in the danger invariant -- a case for which {\sc Dangerzone} is not optimised.
Notably, only 6 of the benchmarks (marked with $*$ in the table) contain 
doomed loop heads.

\begin{figure}
\centering
\small
\begin{tabular}{|l|c|c||c||r|}
\hline
Benchmark & Deep Bugs & Shallow Bugs & \,{\sc CBMC}\, & \,{\sc Dangerzone}\, \\
    \hline
    \hline
\input{danger-table}
    \hline
\end{tabular}

Key: TO = time-out

\caption{Experimental results\label{fig:experiments}}
\end{figure}

\section{Conclusions}

In this paper, we introduced the concept of
danger invariants -- the dual to safety invariants.
Danger invariants summarise sets of traces that are guaranteed to reach an error
state.  This summarisation allows us to find deep bugs without false alarms
and without explicitly unwinding loops.
This new concept promises to
deliver the performance of abstract interpretation together with the
concrete precision of BMC.    We presented a second-order formulation of
danger invariants and used the solver described in~\cite{kalashnikov} to
compute danger invariants for a set of benchmarks.

\bibliography{all}{}
\bibliographystyle{splncs}

\end{document}